\RequirePackage{amsmath}
\documentclass[runningheads]{llncs}

\usepackage{tcolorbox}
\usepackage{etoolbox}

\newif\ifEditMode

 \EditModetrue

\usepackage{listings}
\usepackage{preamble}
\usepackage{acronyms}
\usepackage{aliases}

\usepackage{pifont}% 

\begin{document}

\title{A Public Dataset For the ZKsync Rollup\thanks{MI. Silva and J. Messias contributed equally to this work. Most of this work was performed while the authors were at Matter Labs. The authors also acknowledge the help of Igor Borodin from Matter Labs and his assistance in hosting the dataset.}}

\author{Maria Inês Silva\inst{1,4}
\and Johnnatan Messias\inst{2}
\and Benjamin Livshits\inst{3}}

\authorrunning{MI. Silva, J. Messias, and B. Livshits}

\institute{NOVA Information Management School
\and MPI-SWS
\and Imperial College London
\and Matter Labs
}

\maketitle

\begin{abstract}
  
Despite blockchain data being publicly available, practical challenges and high costs often hinder its effective use by researchers, thus limiting data-driven research and exploration in the blockchain space. This is especially true when it comes to Layer-2 (L2) ecosystems, and ZKsync, in particular. 
To address these issues, we have curated a dataset from~1 year of activity extracted from a ZKsync Era archive node and made it freely available to external parties. We provide details on this  dataset and how it was created, showcase a few example analyses that can be performed with it, and discuss some future research directions.

\end{abstract}

%------------------------------------------------------------------------------
\section{Introduction} \label{sec:introduction}

The blockchain ecosystem is built on the principles of decentralization and transparency. However, a significant challenge remains for end users and non-technical researchers: \stress{blockchain data is not easily accessible}. This challenge hinders the widespread adoption of blockchain technology, which should be more easily accessible.

Currently, blockchain data can be obtained by deploying an archive node in the case of chains based on \gls{EVM}~\cite{ZKsync-doc,BSC,Arbitrum,Optimism}, or a full node for Bitcoin~\cite{BitcoinCore-2024}. However, this requires high bandwidth and a high-speed storage machine to keep the blockchain fully synced. In other words, individual deployment of archive nodes is not a practical solution. Additionally, users can gather the data through \glspl{RPC} providers. However, collecting data this way can be challenging for some non-technical users and researchers and is usually costly.
Alternatively, users can depend on external data sources such as Etherscan, Arbiscan, Dune, ZettaBlock, Nansen, and similar platforms.
Although they may provide an interesting solution, they may prove costly in the long run and do not meet the requirements of specific end-users, particularly those engaged in research that relies on quickly and cheaply accessible data.

We believe that anyone requiring blockchain data should have a straightforward and hassle-free way to access it without worrying about infrastructure or hardware. This data has significant value for research purposes, such as alerting and measurement analysis~\cite{Heimbach@IMC-PBS,Messias@IMC2021}, Airdrop designs and analyzes~\cite{Fan@WWW-Altruistic,messias2023airdrops}, analyzing \gls{MEV}~\cite{gogol2024quantifying,gogol2023cross,Messias@FC2023,qin2022quantifying,torres2021frontrunner,weintraub2022flash}, \gls{AMM}~\cite{gogol2024liquid,qin2021attacking}, and potentially improving the adoption of particular chains such as ZKsync and other \glspl{L2} rollups within the research scope~\cite{messias2024writing}. In addition, it can contribute to the growth of these ecosystems as more researchers become involved in the field.

In a recent talk, a company named Paradigm emphasized that to empower the blockchain community with the necessary data analysis capabilities, we must make blockchain data more accessible~\cite{Reth}. By ensuring that these data are available affordably, quickly, and with minimal effort, we can effectively address the challenge of data unavailability in our ecosystem. They introduced a novel \gls{EVM} blockchain node developed in Rust, known as \stress{Reth}, that can be applied to EVM-compatible blockchains. Paradigm provided an endpoint to their archive node, allowing anyone to make requests to their node. Similar initiatives have been provided by Matter Labs and other companies interested in making blockchain data more accessible~\cite{ZKsync-doc,Arbitrum,Optimism}. Nevertheless, users would need to build high-efficiency code to gather all the necessary data, which could invalidate this process for some users due to network delays or lack of coding skills.

More recently, Paradigm~\cite{Paradigm-data-portal,Paradigm-data-portal-git}, BigQuery~\cite{day_introducing_2019} and other research groups~\cite{Messias@IMC2021,Messias@FC2023} have made blockchain data accessible and available by providing them in an easy-to-download and-load schema. 
In that sense, we followed through and decided to make a ZKsync dataset fully available and accessible to any user or researcher. 
Therefore, we provide~1 year of ZKsync data covering the period of February~14\tsup{th},~2023, (block~1) and March~24\tsup{th},~2024, (block~\num{29710983}). More details of the dataset are available in Section~\ref{sec:data}.

\subsection{Why ZKsync Era Data?}

% Intro to ZKsync Era
Launched in March 2023, ZKsync Era is a \gls{L2} scaling solution for the Ethereum blockchain that utilizes \gls{ZKP} for efficient transaction processing. ZKsync Era is also an \gls{EVM}-based rollup, ensuring compatibility with existing Ethereum smart contracts. In July 2024, ZKsync Era ranked among the top five \gls{ZKP} chains with a \gls{TVL} estimated at~1.23 billion USD~\cite{L2beat}. This scaling solution improves Ethereum by processing transactions in batches using \gls{ZKP}, which helps maintain low transaction fees and encourages user participation.

% Main motivation for ZKsync research
In this sense, rollups are a key strategy to scale the Ethereum ecosystem~\cite{buterin_rollup-centric_2020}. In fact, these \gls{L2} chains have become an important part of the ecosystem in the last few years by pushing new innovations and absorbing a significant portion of user activity. At the same time, there are still many unexplored questions when it comes to these L2 blockchains, and existing research on them is fairly limited. 

Given the growing role of \gls{L2} chains in the Ethereum ecosystem, we believe there is an opportunity for researchers to conduct research and expand our understanding of these L2 blockchains, such as ZKsync. By making a 1-year ZKsync Era data easily available to external groups, we hope to advance ZKsync-related research and knowledge about \gls{L2} chains, more generally.

\subsection{Contributions}

%We back our contributions with the belief that making our ZKsync Era dataset available will significantly benefit research groups interested in studying this emerging area of \gls{L2} chains. Our contributions are summarized as follows.
% Draft

\point{\bf{Public Availability of ZKsync Data.}} To facilitate scientific use of our collected ZKsync Era dataset, we have made it available in a public GitHub repository~\cite{ZKsync-GitHub}.
This consists of a one-year dataset containing information regarding blocks, transactions, receipts, and logs. Dataset details are presented in Section~\ref{sec:data}.

\point{\bf{Facilitating Research and Analysis.}} We demonstrated potential applications of this dataset in research and advocating for the adoption of specific blockchain technologies like ZKsync and \gls{L2} rollups in general in Section~\ref{sec:analyses}.

\point{\bf{Practical Implications for Users and Researchers.}}
We address challenges associated with data gathering via \glspl{RPC} and external platforms like Etherscan, Arbiscan, and others, which may be slow and costly. We also provide an easy-to-download and easy-to-load dataset of ZKsync Era, along with Jupyter notebooks to facilitate this onboarding process~\cite{ZKsync-GitHub}. 
Table~\ref{tab:code} describes the code utilized to load and process our dataset.
All the analyses conducted in this paper can be easily reproduced using our code, which will be available in a GitHub repository. See Sections~\ref{sec:data} and~\ref{sec:analyses} for details.

\subsection{Paper Organization}
This paper is organized as follows. Section~\ref{sec:data} details our dataset and its data schema, together with the necessary background to understand the context of blockchain data in general. Section~\ref{sec:analyses} provides examples of analyses that can be achieved using this dataset, such as regarding transaction fees and gas usage, events derived from transaction logs, token swaps, and more. Furthermore, Section~\ref{sec:future_directions} discusses open problems and future directions for which this dataset could be useful. Finally, we present the conclusion of our work in Section~\ref{sec:conclusion}.

\begin{table*}[t]
\centering
\caption{Description of the code used for analysis in this paper. These notebook files are available in our GitHub repository~\cite{ZKsync-GitHub} in the directory \stress{./zksync-data-dump/notebooks/} and show how to interact and process our ZKsync dataset.}
\resizebox{\textwidth}{!}{
\begin{tabular}{lp{7cm}}
\toprule
\thead{Notebook file} & \thead{Description}     \\
\midrule
    \href{https://github.com/matter-labs/zksync-data-dump/blob/main/notebooks/01-zksync-data.ipynb}{01-zksync-data.ipynb} & \scriptsize{It computes the basic statistics of the dataset and provides analyses used in Section~\ref{sec:data}. We use four main sources of data: \stress{blocks}, \stress{transactions}, \stress{transaction receipts}, and \stress{logs}}.\\
    \href{https://github.com/matter-labs/zksync-data-dump/blob/main/notebooks/02-data-exploration-fees.ipynb}{02-data-exploration-fees.ipynb} & \scriptsize{It analyses gas usage and transaction fees for ZKsync used in Section~\ref{subsec:gas-and-fees}. We use two main sources of data: \stress{blocks} and \stress{transaction receipts}}.\\
    \href{https://github.com/matter-labs/zksync-data-dump/blob/main/notebooks/03-data-exploration-contracts.ipynb}{03-data-exploration-contracts.ipynb} & \scriptsize{It analyzes the contract deployment and events triggered on ZKsync described in Section~\ref{subsec:contracts-events}. We use one main source of data: \stress{transaction logs} but also load \stress{blocks} data to extract timestamps information.} \\
    \href{https://github.com/matter-labs/zksync-data-dump/blob/main/notebooks/04-data-exploration-swaps.ipynb}{04-data-exploration-swaps.ipynb} & \scriptsize{It analyzes the swap events on ZKsync described in Section~\ref{subsec:swaps}. We use one main source of data: \stress{transaction logs} but also load \stress{blocks} data to extract timestamps.} \\
\bottomrule
\end{tabular}}
\label{tab:code}
\end{table*}

\section{Data Schema and Processing} \label{sec:data}

In this section, we present our ZKsync dataset in detail. The dataset covers the period of February~14\tsup{th},~2023, (block number~1) and March 24\tsup{th},~2024, (block number~\num{29710983}) corresponding to~1 year of data. It contains~\num{327174035} transactions and~\num{1631772} contracts deployed during this time period which triggered~\num{2044221151} events on-chain. Transactions were issued by~\num{7322502} unique users. This dataset enables researchers or blockchain enthusiasts to explore all the activities that occurred in the ZKsync Era since its deployment. Table~\ref{tab:dataset} summarizes our ZKsync Era dataset.

\begin{table*}[t]
\centering
\caption{Description of our 1-year ZKsync dataset.}
\resizebox{\textwidth}{!}{%
\begin{tabular}{lccrrrrr}
\toprule
\thead{Chain} & \thead{Start date} & \thead{End date} & \thead{\# of Blocks}  &
\thead{\# of Transactions} & \thead{\# of Issuers}  & \thead{\# of Contracts} & \thead{\# of Logs}     \\
\midrule
    ZKsync Era & February 14\tsup{th}, 2023 & March 24\tsup{th}, 2024  & \num{29710983} & \num{327174035}  & \num{7322502} & \num{1631772} & \num{2044221151} \\
\bottomrule
\end{tabular}}
\label{tab:dataset}
\end{table*}

We gathered our dataset from a ZKsync Era archive node as raw data. This data consists of all the information regarding blocks, transactions, receipts, and logs. Then, we conducted a pre-processing step to allow anyone to use the dataset. This step consists of formatting the data in a parquet format that can be easily accessible through well-known libraries available in Python, for example, Pandas~\cite{Pandas} and Polars~\cite{Polars}. Due to the high volume of data (i.e., around~200~GB), we focus on using Polars for better processing, memory management, and \stress{Lazy evaluation}~\cite{johnsson1984efficient,Polars-Lazy}, allowing straightforward data processing on a local laptop. Next, we discuss each of the different data types separately.

\subsection{Blocks} 
\begin{figure}[t]
\centering
\includegraphics[width=\onecolgrid]{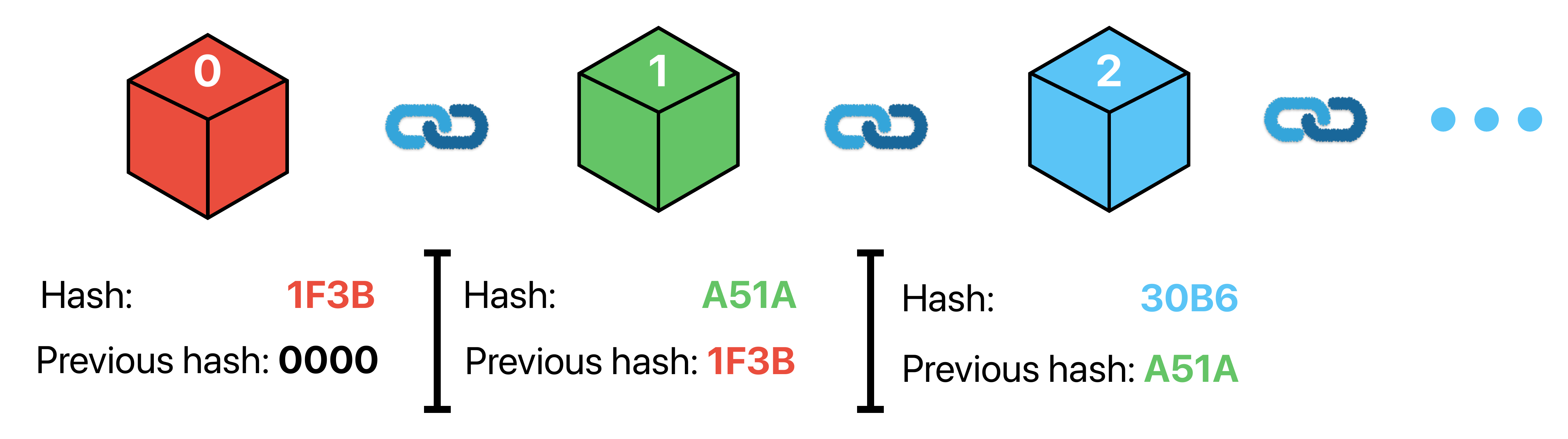}
\figcap{Illustration of blocks connected to each other forming the blockchain.}
\label{fig:blockchain}
\end{figure}

Blocks are sequential units of data within a blockchain, each identified by a unique hash. They contain a list of transactions, metadata such as timestamps, and the hash of the previous block (\stress{parentHash}), which links them in a chain back to the genesis block (block number 0) as shown in Figure~\ref{fig:blockchain}. This chain of blocks forms the blockchain. Blocks ensure transaction security, network consensus, and efficient data storage and processing within blockchain networks.
In our dataset, blocks should be sorted according to the attribute \stress{block number} to maintain the correct order of the blocks on the blockchain.

\subsection{Transactions} Transactions are digital interactions that involve the transfer of assets, the recording of data, or the execution of smart contracts between parties on blockchains like those based on \gls{EVM}. Each transaction is initiated by a user, authenticated through cryptographic signatures, and sent to a decentralized network of nodes for validation. Once verified, transactions are grouped into blocks and added to the blockchain via a consensus mechanism, ensuring that they are secure, immutable, transparent, and free of intermediaries, forming the core of the blockchain system.

In rollups, such as ZKsync, transactions are aggregated and processed off the underlying blockchain (e.g., Ethereum, a~\gls{L1}) to enhance scalability and reduce costs. Rollups bundle multiple transactions into a single batch, which is then submitted to the underlying blockchain as one transaction. This method reduces the load on the underlying chain while ensuring transaction security and finality through cryptographic proofs, such as \gls{ZKP} used by ZKsync or validity checks. By processing transactions off-chain and periodically committing the results to the underlying chain, rollups improve throughput and efficiency without compromising the security and decentralization of the blockchain.

Transactions are identified by a unique transaction hash. When issuing a transaction, the user needs to specify parameters such as the recipient address (which can also be a smart contract and the functions the user wants to call), the number of tokens to transfer, the gas price, and the gas limit. The gas price represents the fee the user is willing to pay per unit of gas, while the gas limit is the maximum amount of gas the user is willing to consume for the transaction, a mechanism introduced to prevent infinite loops or excessive resource consumption.

In our dataset, transactions should be sorted by the \stress{blockNumber} and \stress{transactionIndex} attributes to maintain the correct order of the transactions on the blockchain.

\subsection{Transactions Receipts}

Besides the transaction data, our dataset also contains a table of transaction receipts.
These transaction receipts provide a comprehensive summary of the outcome and effects of a transaction once it is processed and included in a block. They include the transaction hash, block number, and block hash to identify and verify the transaction, along with the sender (\stress{from}) and recipient (\stress{to}) addresses. 
The receipts also detail the cumulative gas used by the transaction and all preceding transactions in the block, the actual gas used by the specific transaction, and the final gas price paid by the user. By multiplying the actual gas used by the gas price, we have the actual transaction fee the user paid.

Additional information includes \stress{logs} for event logging (discussed next), the transaction status (success or failure), and the effective gas price paid. These receipts are crucial for users and developers to understand, audit, and interact with transactions and smart contracts on the blockchain. For example, they provide essential elements for analyzing, monitoring, and verifying transaction fees spent by users. 

Similarly to transactions, in our dataset, transaction receipt data should be sorted by attributes \stress{blockNumber} and \stress{transactionIndex} to maintain the correct order of the transactions on the blockchain.

\subsection{Transactions Logs}

In this section, we discuss the attributes of the transaction logs data in the ZKsync dataset in detail. Transaction logs are systematic records of events generated during the execution of transactions, particularly in interactions involving smart contracts. Each log entry contains \stress{log index}, \stress{data}, and \stress{topics}, crucial to identifying and categorizing specific events such as token transfers, approvals, swaps, minting, and voting. 

These logs are emitted using the \stress{emit} keyword within the smart contract code and play a pivotal role in monitoring activities, triggering actions within decentralized applications, and enabling event-driven programming. For instance, \glspl{DEX} emit events upon trade executions, enabling user interfaces to update displays with current trade information.
These logs are stored in transaction receipts, offering a gas-efficient method to capture transient event data without permanently altering the blockchain's state. Among the vast array of data accessible on \gls{EVM}-based blockchains, transaction logs stand out as crucial sources of information for researchers, developers, and users. They facilitate analyses of various token transfer patterns and support blockchain analysis research, which is the focus of our contributions.

In our dataset, transaction logs should be sorted by the \stress{blockNumber}, \stress{transactionIndex}, and \stress{logIndex} attributes to maintain the correct order in which they are stored on the blockchain. This is particularly important when analyzing the different states of a blockchain before and after the execution of a transaction that triggers a smart contract function.

\subsubsection{Topics Attributes.}

The interpretation of the \stress{topics} attributes (\stress{topics$_0$}, \stress{topics$_1$}, \stress{topics$_2$}, and \stress{topics$_3$}) depends on the implementation details of the invoked function within a smart contract. Typically, \stress{topics$_0$} represents the event name, while subsequent topics represent indexed parameters of the event. The ``data'' attribute contains non-indexed event parameters. For example, in the context of a token transfer event, \stress{topics$_0$} might indicate the event name \stress{Transfer}, \stress{topics$_1$} and \stress{topics$_2$} could respectively denote sender and receiver addresses, and \stress{data} would typically represent the number of tokens transferred.

\subsubsection{Hashing and Signatures.}

\stress{topics$_0$} corresponds to the hashed function signature using \stress{keccak256}~\cite{bertoni2009keccak}. This signature consists of the function name followed by its parameter types. For example, the signature of a typical \stress{Transfer} event is \stress{Transfer(address,address,uint256)}. After hashing it with keccak256, the result becomes \stress{0xddf2 $\cdots$ b3ef},\footnote{We shortened \stress{topics$_0$} to \stress{0xddf252ad $\cdots$ f523b3ef} for better visualization in the paper.} which is the \stress{topics$_0$}. Below is a Python code snippet demonstrating how to verify if a given signature matches \stress{topics$_0$}:

\footnotesize
\begin{lstlisting}[language=Python]
import web3
def check_sig(sig, topics_0):
    return web3.Web3.keccak(text=sig).hex() == topics_0
\end{lstlisting}

\subsubsection{Event Mapping.}

We provide a mapping of the top 90 most frequently invoked events within the ZKsync dataset in our GitHub repository under \stress{./src/utils.py\#events\_dict}.
This mapping facilitates the parsing of the majority of events in our dataset. The mapping is structured as a dictionary where the topics$_0$ hex value serves as the key, and the corresponding value is a dictionary containing the parsed event name and its function signature. For instance, the Transfer event is represented as follows within the map:

\footnotesize
\begin{lstlisting}[language=Python]
events_dict["0xddf252ad1be2c89b69c2b068fc378daa952ba7f163c4a11628f55a4df523b3ef"] = {
    "name": "Transfer",
    "signature": "Transfer(address,address,uint256)"}
\end{lstlisting}

\subsection{L2 to L1 Logs}

\gls{L2} to \gls{L1} logs are messages emitted by the ZKsync \gls{L2} network and transmitted to the Ethereum \gls{L1} network. They are essential for maintaining communication between the two layers, ensuring the security and integrity of transactions and data transfers. In ZKsync, the \gls{L1} smart contract verifies these communications by checking the messages alongside the \gls{ZKP}. The only \stress{provable} part of the communication from \gls{L2} to \gls{L1} is the native \gls{L2} to \gls{L1} logs emitted by the \gls{VM}. These logs can be generated using the \stress{to\_l1 opcode}~\cite{ZKsync-doc-contracts-bootloader}. We refer the reader to the ZKsync documentation for more details~\cite{ZKsync-doc-l2-to-l1-logs}.
\section{Example Analyses} \label{sec:analyses}

This section demonstrates some analyses that can be performed using our ZKsync dataset. Each subsection concentrates on a specific topic and utilizes their respective data (e.g., blocks, transactions, receipts, and logs). In this paper, we consider all activity that starts with block~\num{561367}, the first block on April 1\tsup{st}, until block~\num{29710983}, the last block in the ZKsync dataset.

This section should serve as a starting point for other researchers wishing to use this dataset for their research. The code for generating this analysis is discussed in Table~\ref{tab:code} and will be available in a GitHub repository.

\subsection{Gas Usage and Transaction Fees} \label{subsec:gas-and-fees}

To start, we look into transactions, gas usage, and fees. All results are generated from two main sources of data, namely blocks and transaction receipts. Note that we are using the transaction receipts instead of the transaction data because the receipts are the source of the actual units of gas used and the final transaction fee paid by users.

\begin{figure}[t]
\centering
\includegraphics[width=\onecolgrid]{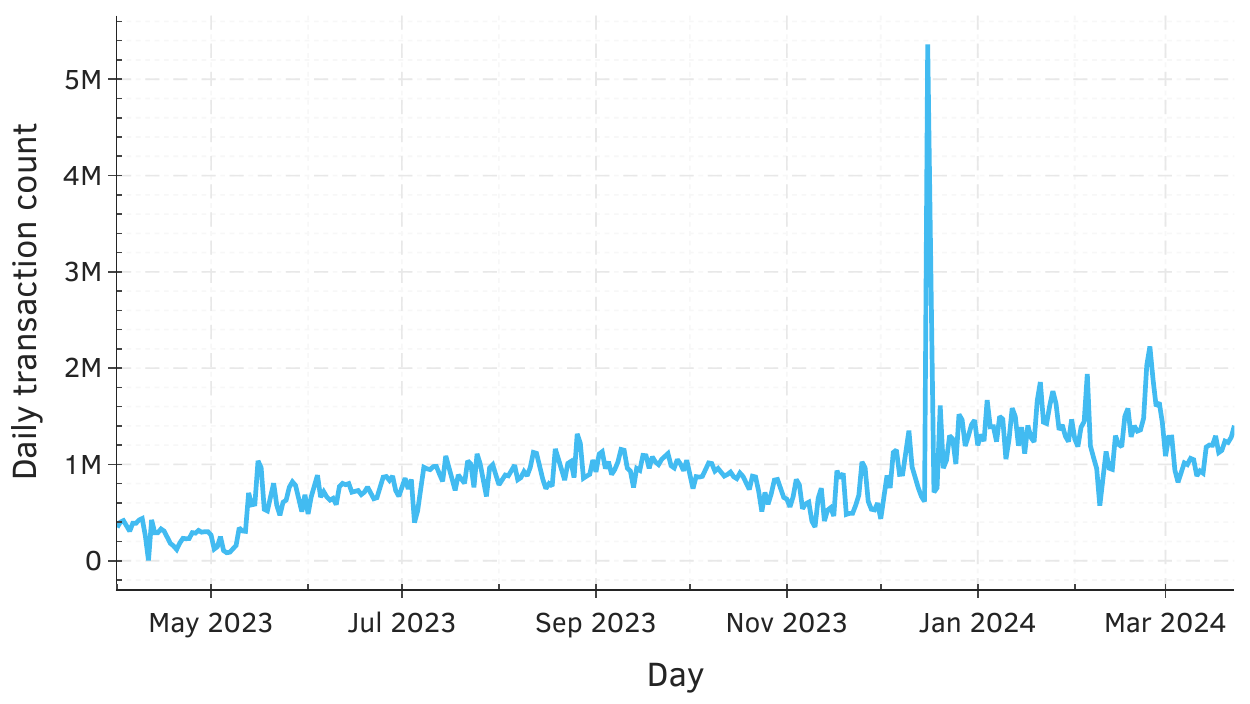}
\caption{Total transactions executed per day.}
\label{fig:total-tx-day}
\end{figure}

Figure~\ref{fig:total-tx-day} shows the daily transactions executed in ZKsync Era over the period analyzed. The network processed an average of \num{905194} transactions daily, with a significant spike in December~2023 that accounts for \num{5362921} transactions in a single day. This spike was due to a boom in inscriptions, which became very popular for memecoin traders around this time. Messias~\ea~\cite{messias2024writing} did a comprehensive review of this phenomenon across various rollups.  
Since the spike stabilized, we have seen a slight increase in daily transactions, which are now hovering around~1.2 million transactions per day.

\begin{figure}[t]
\centering
\begin{subfigure}{\twocolgrid}
    \centering
    \includegraphics[width=\twocolgrid]{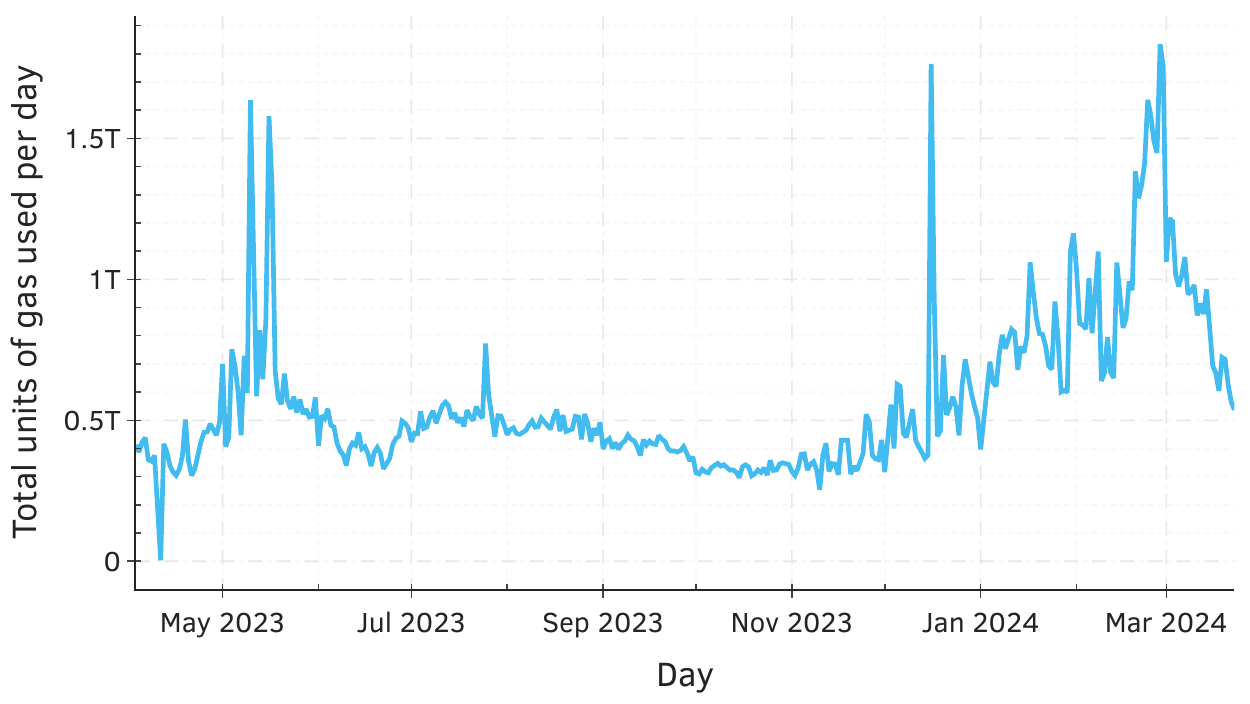}
    \caption{Total gas units.}
    \label{fig:daily-total-gas-used}
\end{subfigure}
\begin{subfigure}{\twocolgrid}
    \centering
    \includegraphics[width=\twocolgrid]{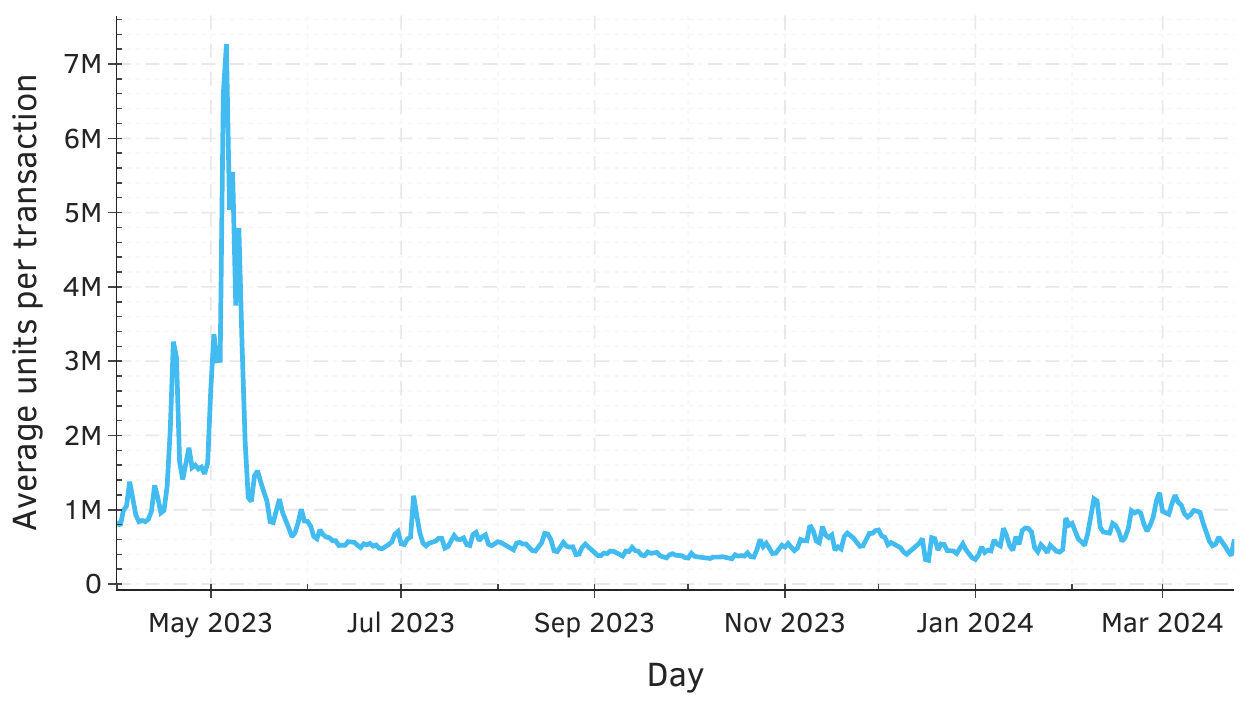}
    \caption{Average gas unit.}
    \label{fig:avg-gas-used-tx}
\end{subfigure}
\caption{Comparison of gas unit usage over time: (a) Total gas units used per day; and (b) Average gas units used per transaction.}
\label{fig:gas-units}
\end{figure}

Considering gas usage, Figure~\ref{fig:daily-total-gas-used} illustrates the total number of gas units used each day in the ZKsync Era. On average, the network has processed \num{567111979658} gas units per day in the last year. However, gas usage has been relatively volatile. Particularly, it has experienced three significant spikes of more than~1.5 trillion gas units per day, which we detail next.

\paraib{May 2023.} 30\% of all gas used during the month of May~2023 can be attributed to the \stress{zkApes airdrop}. The address receiving the most gas units was the \stress{SyncSwap router} contract, which is used to execute swaps in the largest \gls{DEX} in ZKsync Era. During this time, we see more than~10\% 

\paraib{December 2023.} This was caused by the inscriptions boom we discussed before~\cite{messias2024writing}.

\paraib{March 2024.} The gas utilization during this spike is much more distributed among transaction receivers, with the top receivers being \gls{DEX} routers, such as the \stress{SyncSwap V2 router} and the \stress{Mute.io router} (which recently re-branded to Koi), and token contracts, such as \stress{USDC.e} and \stress{SOUL}. This dispersed distribution of gas usage coupled with the observed steady growth in both gas usage and transaction count up to the spike suggests that it is due to the increased trading activity on ZKsync Era.

We can further contextualize gas utilization by looking at Figure~\ref{fig:avg-gas-used-tx}, which shows the average number of gas units used per transaction. We see that in May~2023, transactions were using on average a significantly higher number of gas units (reaching \num{7265323} units per transaction) compared to the average over the~1-year period (\num{784149} units).

\begin{figure}[t]
\centering
\includegraphics[width=\onecolgrid]{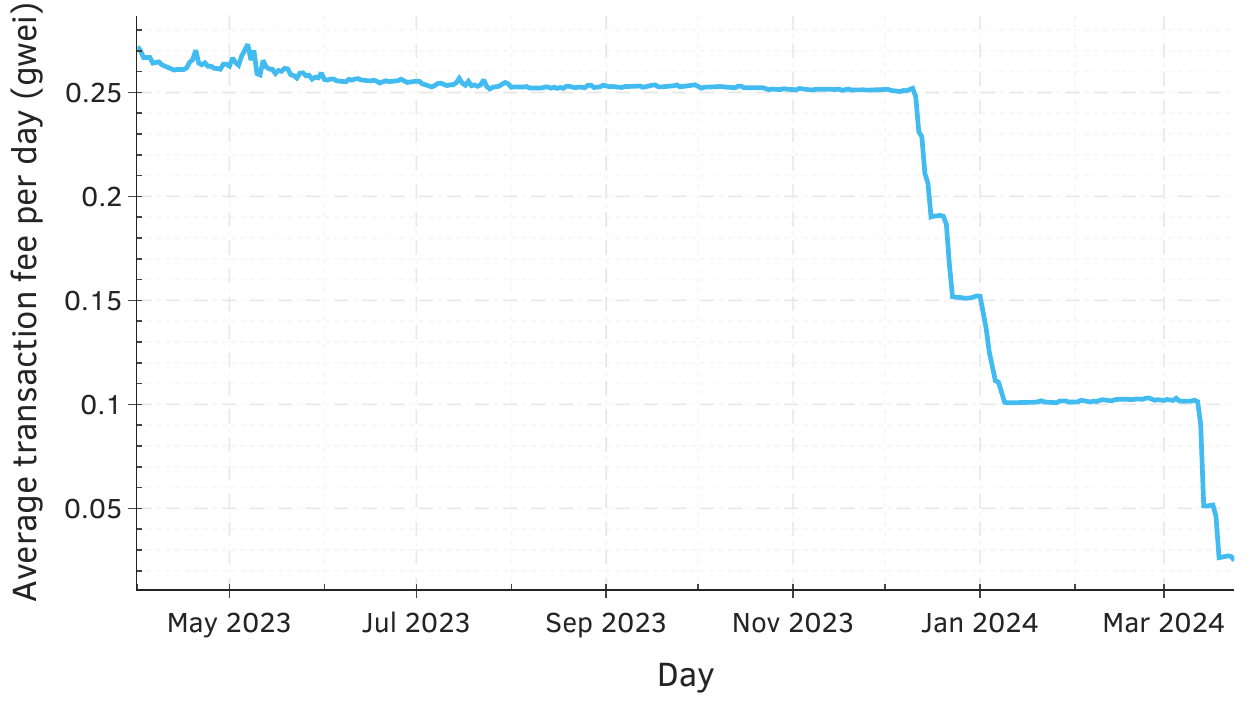}
\caption{Average transaction fee in gwei per unit of gas by day.}
\label{fig:avg-tx-fee-day}
\end{figure}

Finally, we can look at the transaction fees. Figure~\ref{fig:avg-tx-fee-day} shows the average transaction fees over time. In other words, this is the cost that users have paid, on average, to submit one unit of gas to ZKsync Era. Transaction fees were fairly stable during 2023, with an average of~0.25 gwei per gas unit. 

However, in late 2023, a key upgrade led to a significant decrease in ZKsync transaction fees. Concretely, the implementation of the new prover, Boojum~\cite{Boojum}, marked a significantly reduced hardware requirement to run a \gls{ZK} prover and thus allowed a reduction of transaction fees to~0.1 gwei per unit of gas.

Then, in March 2024, the Dencun upgrade was implemented and deployed on the Ethereum mainnet. This upgrade, brought in EIP-4844, introduced a new type of transaction that can store ``blobs'' of data in the beacon node for~14 days~\cite{EIP-4844}. Blobs have their independent fee model and submitting data as blobs is much cheaper than the previously used call-data~\cite{Godbole@Coindesk}. ZKsync Era was one of the first rollups using blobs to publish the data related to its state changes, thus further reducing transaction fees to~0.025 gwei per unit of gas.

\subsection{Events and Contract Deployments} \label{subsec:contracts-events}

Events are an important source of additional data about activity on \gls{EVM} chains. They correspond to data emitted by smart contracts and stored on-chain. Next, we analyze this data and report on some specific events. Recall that events can be obtained from the logs data in the ZKsync dataset as discussed in Section~\ref{sec:data}.

\begin{figure}[t]
\centering
\includegraphics[width=\onecolgrid]{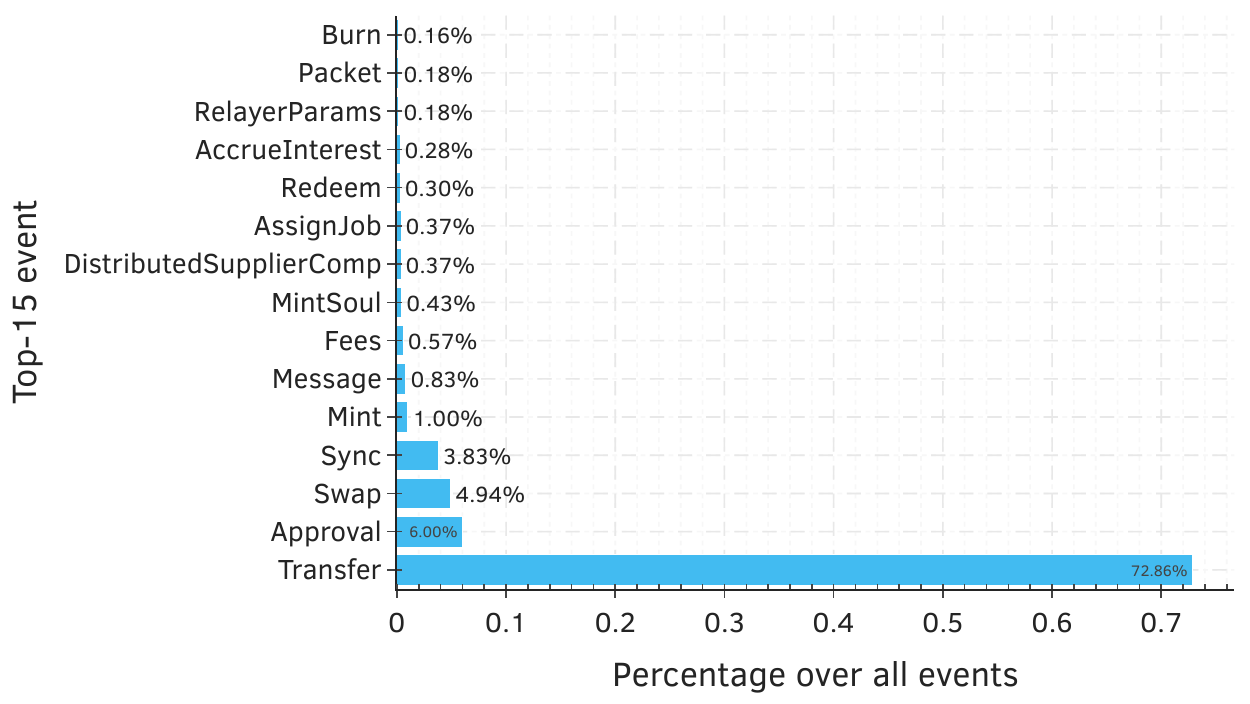}
\caption{Top 15 event types with the highest number of events emitted between April~1\tsup{st}, 2023 and March~24\tsup{th},~2024, and their percentage over all events emitted. Percentages \stress{include} ETH transfers generated from transaction fee payments.}
\label{fig:top-events}
\end{figure}

Figure~\ref{fig:top-events} shows the top 15 event types with the most emitted events during the period under analysis. Different contracts may emit the same event type, so what distinguishes them is the event's function signature and, thus, its respective hash.

There is a significant overrepresentation of the top~4 event types, with \stress{Transfer} events being by far the most significant type (70.9\% of all events are Transfer events). This is not unexpected as this event is emitted every time an ERC-20 token is transferred between two addresses in ZKsync Era. This occurs in simple token transfers and other standard contract operations such as swaps in \glspl{DEX}.

Before looking at \stress{Transfer} events in more detail, we should highlight a particularity in how ZKsync Era implements transaction fee collection. In this chain, every transaction generates two additional ETH transfers~---~one for the initial payment of the transaction fees from the transaction submitter and another with a transaction fee refund (after all L1 and proving costs are accounted for). These transaction fee transfers always appear as ETH transfers from or to the address \stress{\href{https://era.zksync.network/address/0x0000000000000000000000000000000000008001}{0x8001}}\footnote{We shortened this address for better visualization in the paper.} and thus generate \stress{Transfer} events.
These specific events account for~38.2\% of all events emitted, and if we exclude them entirely from the event dataset, Transfer events only account for~56.1\% of these filtered events.

\begin{figure}[t]
\centering
\begin{subfigure}{\twocolgrid}
    \centering
    \includegraphics[width=\twocolgrid]{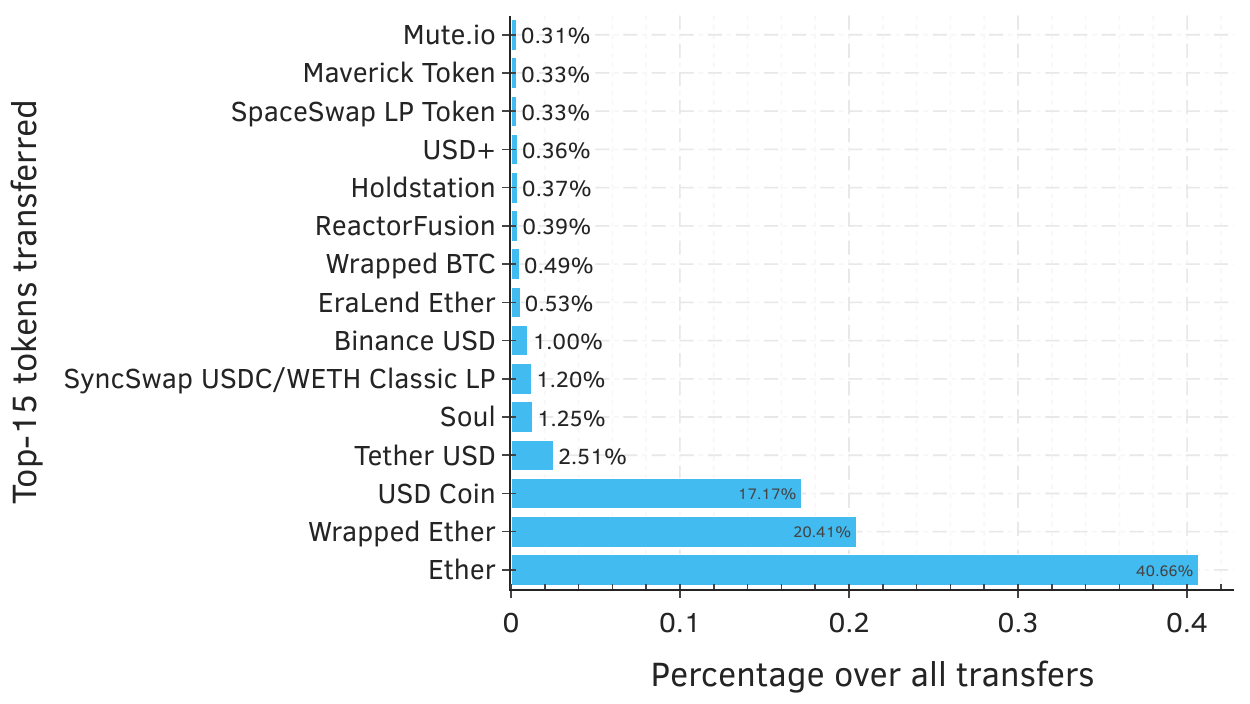}
    \caption{Top~15 ERC-20 tokens with the highest number of \stress{Transfer} events emitted.}
    \label{fig:top-tokens}
\end{subfigure}
\begin{subfigure}{\twocolgrid}
    \centering
    \includegraphics[width=\twocolgrid]{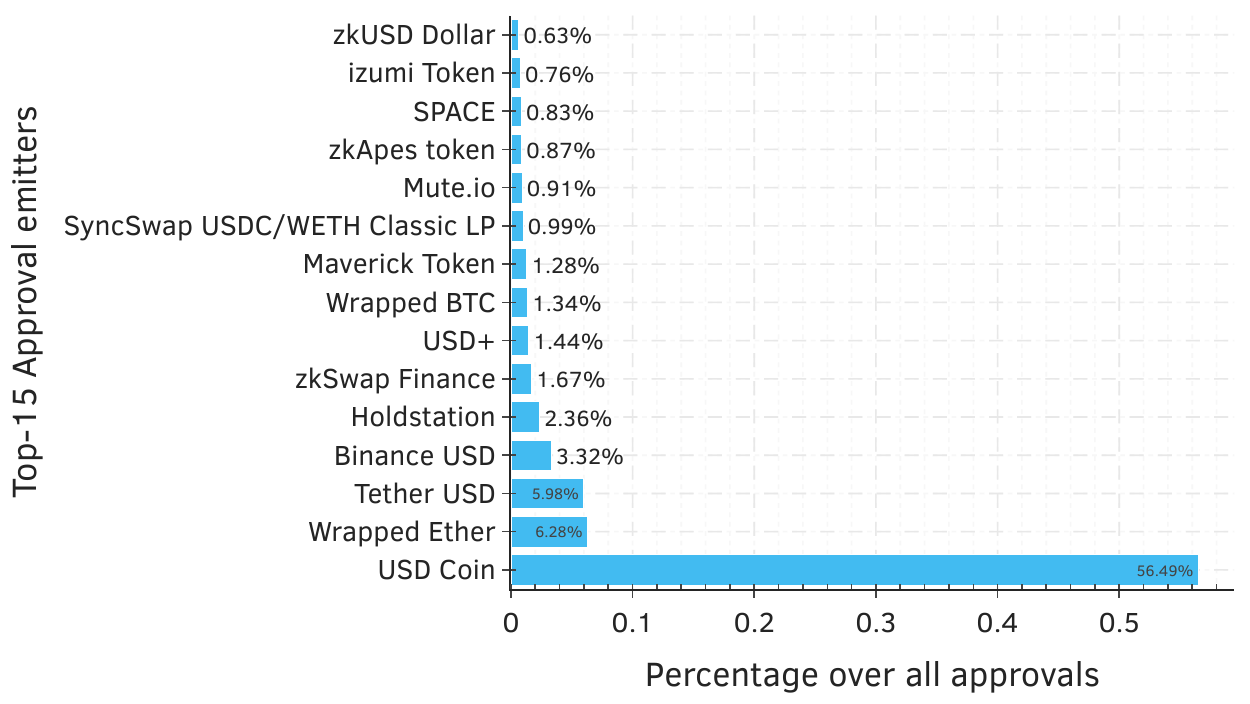}
    \caption{Top 15 contracts with the highest number of \stress{Approval} events emitted.}
    \label{fig:top-approval-emitters}
\end{subfigure}
\caption{Percentage of the top~15 contracts with the highest Transfer and Approval events emitted between April~1\tsup{st},~2023 and March~24\tsup{th},~2024: (a) Transfer events, where percentages \stress{exclude} ETH transfers generated from transaction fee payments; and (b) Approval events.}
\label{fig:top-events-transfer-and-approval}
\end{figure}

After filtering the Transfer events generated by fee management ETH transfers, we can explore the top tokens transferred by examining the contract address that emitted the event. Figure~\ref{fig:top-tokens} displays the~15 tokens most frequently transferred during the period.
Native ETH, Wrapped ETH, and Bridged USDC (USDC.e) are the most transfers, accounting collectively for~78.4\%. We can also see some other stablecoins, such as Tether USD or Binance USD, and \gls{LP} tokens associated with \glspl{DEX} on ZKsync Era, such as SyncSwap and SpaceSwap.

\stress{Approvals} are the second-largest event type, with~6\% of events emitted. Similarly to the Transfer events, we can see which contracts emit these events. Figure~\ref{fig:top-approval-emitters} shows these top~15 emitters.

The top emitters are also ERC-20 token contracts, and these events represent an owner ``approving'' a spender to transfer a predefined amount of tokens they hold. This is common in bridged assets, for example, as is the case of Bridged USDC (USDC.e) and Wrapped ETH, the top~2 emitters.

\stress{Swaps} and \stress{Syncs} are the third and fourth most frequently emitted event types, accounting for~4.9\% and~3.8\% of all events emitted, respectively. These events are a key source of data for analyzing \glspl{DEX}. However, we will explore these more in-depth in the next subsection.

Finally, we examine contract deployments. Every time a contract is deployed, a specific event type is emitted, which allows us to easily track this metric. Although this event type is not among the 15 most emitted event types, it is still an important metric for network activity. In ZKsync, the contract deployment event is named \stress{ContractDeployed}.

\begin{figure}[t]
\centering
\begin{subfigure}{\twocolgrid}
    \centering
    \includegraphics[width=\twocolgrid]{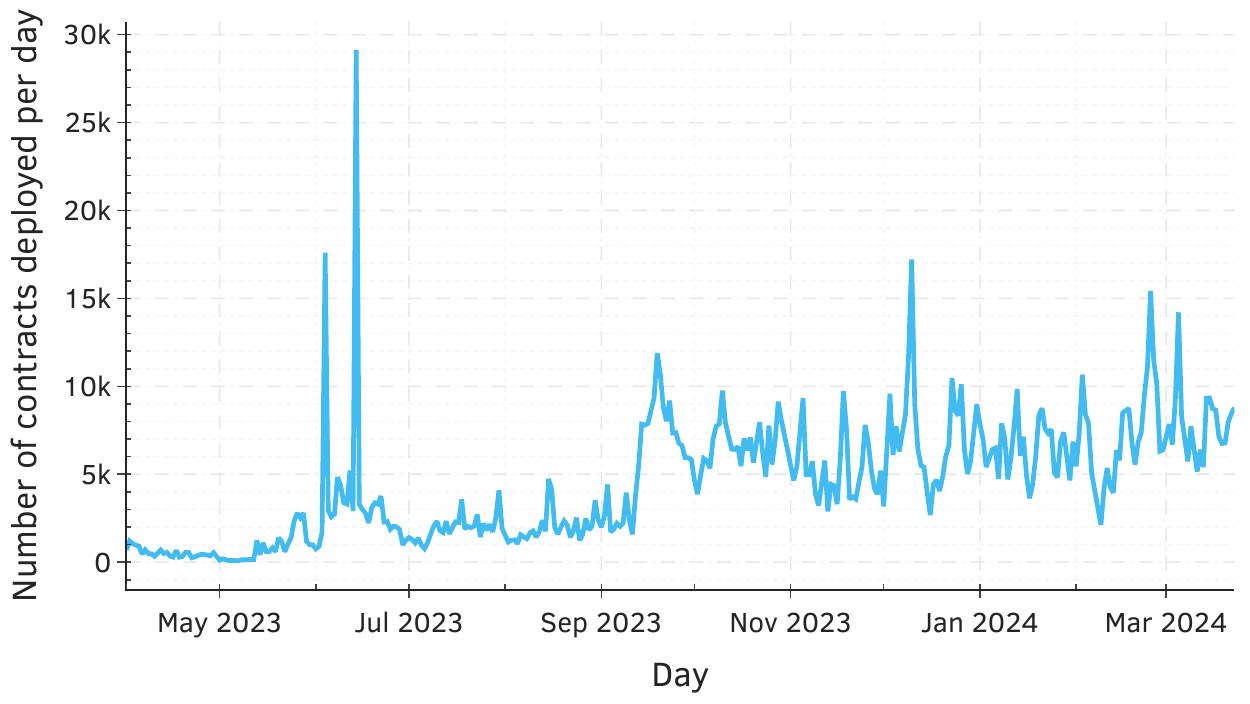}
    \caption{Number of contracts deployed.}
    \label{fig:contracts-deployed-day}
\end{subfigure}
\begin{subfigure}{\twocolgrid}
    \centering
    \includegraphics[width=\twocolgrid]{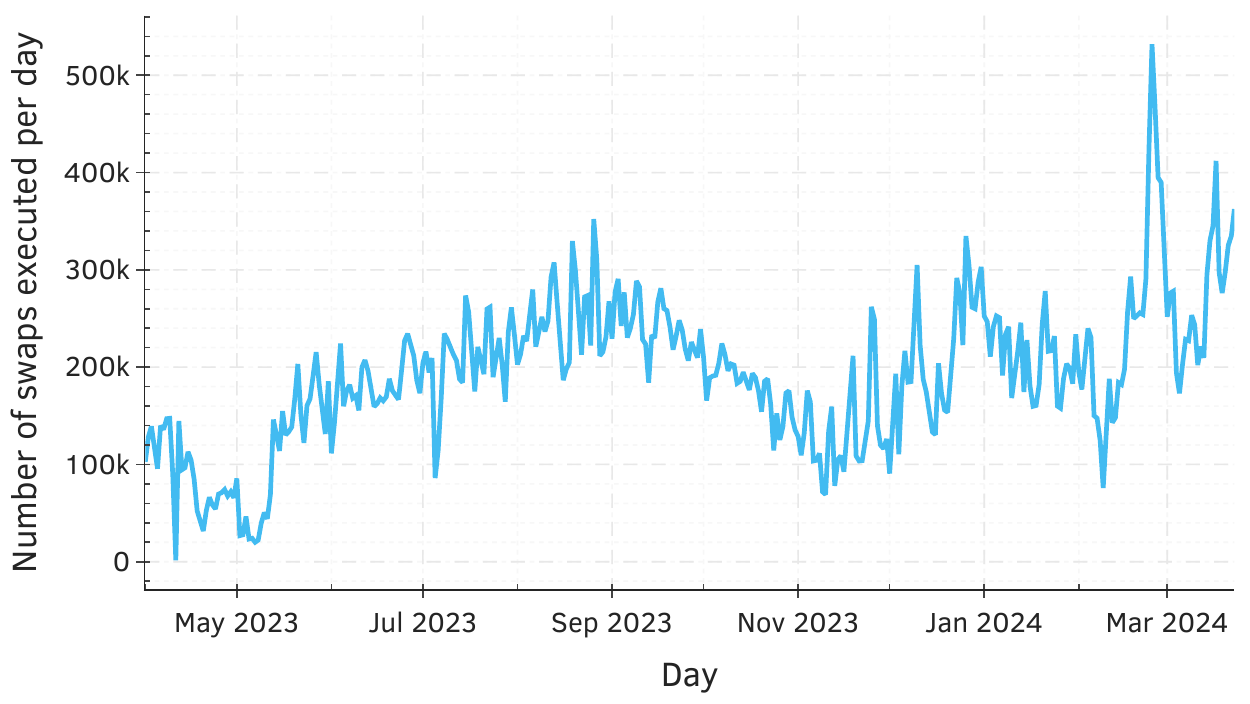}
    \caption{Number of swap events emitted.}
    \label{fig:swaps-per-day}
\end{subfigure}
\caption{Daily total number of contracts deployed (in a) and daily number of swap events emitted (in b) on ZKsync Era.}
\label{fig:contracts-swaps}
\end{figure}

Figure~\ref{fig:contracts-deployed-day} shows the number of contracts deployed on ZKsync each day. Before September~2023, developers averaged \num{2510} contract deployments per day. However, there were two major spikes during this time, the first reaching \num{29114} and the second reaching \num{17602} contract deployments in a single day. Then, after September~2023, contract deployments increased to a daily average of~\num{6672}.

\subsection{Swaps} \label{subsec:swaps}

After a high-level look at these events, we now focus on a specific event relevant to understanding activity on \glspl{DEX}~---~the \stress{Swap} event. Recall that swap events account for~4.9\% of all events emitted on ZKsync.

Swap events are emitted every time a successful swap is performed in a \gls{DEX}. These events contain information about the contract that emits the event, the amounts of each token being traded, and the wallets involved in the trade. 

Figure~\ref{fig:swaps-per-day} shows the number of swap events emitted each day during the analyzed period, which is equivalent to the number of swaps performed each day on ZKsync \glspl{DEX}. We have a long-term trend of increasing the number of daily swaps, with a peak around March 2024 (which reached \num{531819} swaps in a single day). During this year, users performed an average of \num{192009} swaps per day.

We also see which \gls{LP} contract was involved in the swap by looking at the contract address that emitted the event. Depending on the \glspl{DEX} protocol, this may be the actual liquidity pool involved in the swap (e.g., SyncSwap and Koi) or a generic \gls{LP} contract managing all the pools in the \gls{DEX} (e.g., SpaceSwap and PancakeSwap). To understand which tokens were being traded in these generic \gls{LP} contracts, we would have to process the transfer events emitted in the same transaction of these Swap events. 

\begin{figure}[t]
\centering
\includegraphics[width=\onecolgrid]{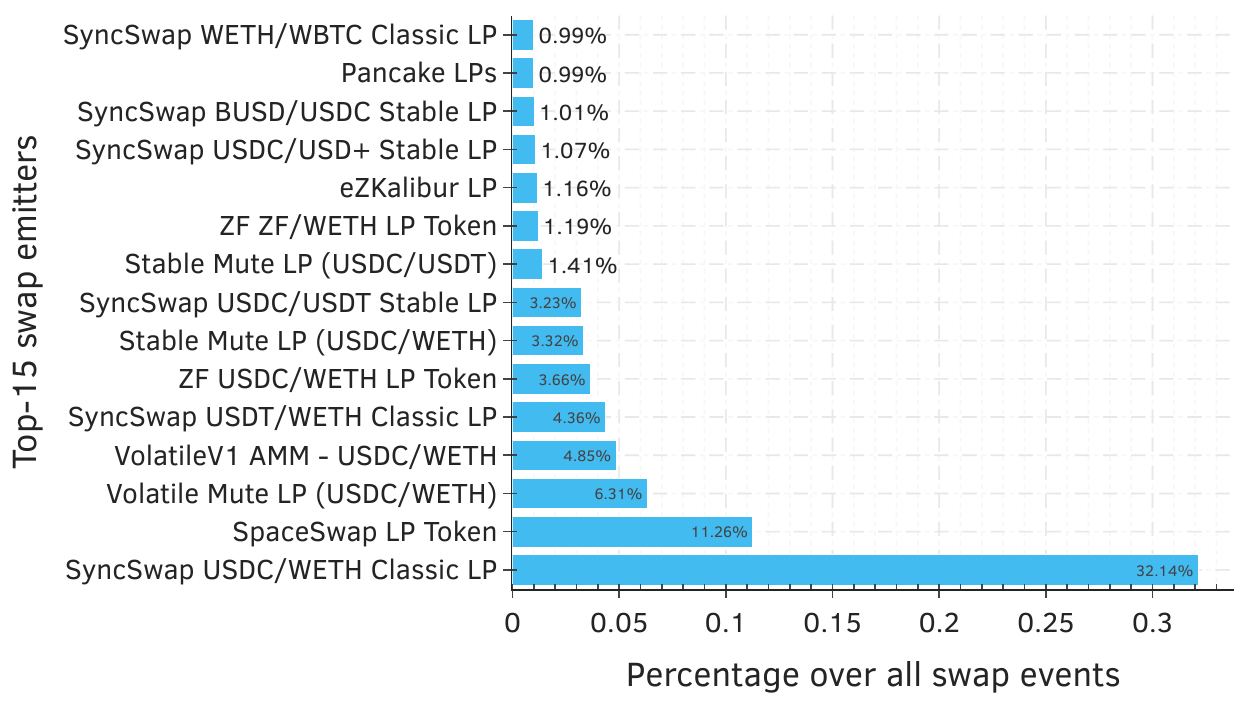}
\caption{Top~15 contracts emitting the most Swap events between April 1\tsup{st}, 2023 and March 24\tsup{th}, 2024, and their percentage over all Swap events emitted.}
\label{fig:top-swap-emitters}
\end{figure}

Figure~\ref{fig:top-swap-emitters} shows the top 15 contracts that emitted the most swaps even during the period under analysis. The largest emitter is, by a significant margin, the USDC/WETH pool on SyncSwap, which accounts for almost a third of all emitted Swap events. All combined swap events from SpaceSwap represent~11\%  of all swaps, followed by two USDC/WETH pools from other \gls{DEX}.

\begin{figure}[t]
\centering
\includegraphics[width=\onecolgrid]{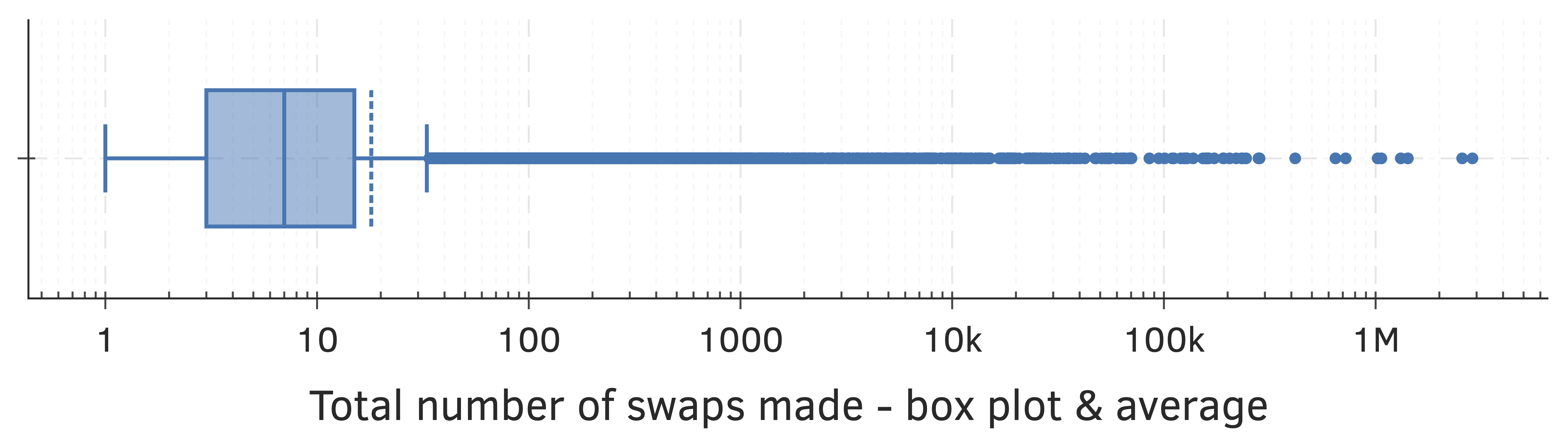}
\caption{Distribution of swaps per unique wallet addresses.}
\label{fig:swaps-by-swapper-dist}
\end{figure}

Finally, we can use the ``receivers'' field to analyze swappers. In other words, this field provides the wallet address receiving the swapped token. Figure~\ref{fig:swaps-by-swapper-dist} displays the distribution of swaps made by unique wallet addresses. 

We note that most users performed less than~50 swaps during the period under analysis. Concretely, the percentile~95\% of this distribution is~42 swap events. However, the distribution has a significant skew to the right, with a few addresses generating a large number of swaps in the year examined. These large ``traders'' are usually routers and other protocols that interact with \glspl{DEX}; thus, they represent many end-users. Examples include the top swappers, which are the \stress{Mute.io router}, the \stress{SpaceFi router}, and the \stress{Odos V2 router}, respectively.

\section{Future Directions} \label{sec:future_directions}

Among the data analysis presented in this paper, our ZKsync dataset can be applied to different studies. We list some of them below, where this dataset can be valuable.

\paraib{MEV and Arbitrage.}
\gls{MEV} and arbitrage have been extensively studied in \gls{L1} blockchains~\cite{Messias@FC2023,qin2022quantifying,qin2021attacking,torres2021frontrunner,weintraub2022flash}. However, only recently have new studies shifted their focus to \glspl{L2}~\cite{bagourd2023quantifying,gogol2024crossrollupmevnonatomicarbitrage,gogol2024quantifying,torres2024rolling}, including ZKsync Era. Our dataset can contribute to this type of research by enabling further analysis of \gls{MEV} on ZKsync Era. For example, one type of \gls{MEV} known as \textit{backrunning} involves arbitrageurs ensuring their transactions are included immediately after a target transaction. This can be particularly useful in scenarios like liquidation-MEV, where arbitrageurs take advantage of opportunities that arise right after an oracle update~\cite{Messias@FC2023,qin2022quantifying,weintraub2022flash}. Another form of arbitrage worth analyzing is CEX-DEX \gls{MEV}, where arbitrageurs exploit price deviations between \gls{CEX} and \gls{DEX} platforms~\cite{Heimbach-SP-Arbitrage}. Additionally, studying cross-rollup \gls{MEV} could provide insights into opportunities where arbitrageurs benefit from differences across two or more rollups~\cite{gogol2024crossrollupmevnonatomicarbitrage}.

\paraib{Analysis of user activity on chain.}
Analyzing user activity in the ZKsync chain provides valuable insights into user behavior. This includes examining the transactions issued by users and their interactions with smart contracts. Such an analysis can also help identify airdrop farmers who create multiple accounts, known as Sybils, although this task can be challenging~\cite{Fan@WWW-Altruistic,messias2023airdrops}. Another important area of analysis is measuring the impact of users in social media networks driving activities on the blockchain. For example, examining the recent boom in inscriptions in \glspl{L2} chains can reveal how social networks can influence blockchain activity~\cite{messias2024writing,Omena@Poster}. Additionally, since our dataset includes all event logs, it allows for the analysis of decentralized governance of protocols deployed on ZKsync Era. This involves studying proposals to amend smart contracts and their voting processes. By filtering data related to governance contracts, such as how each user voted and the distribution of these governance tokens among users, it can shed light on the implications of token concentration on decentralized governance, raising concerns about fairness~\cite{Messias-governance@ArXiv,sharma2023unpacking}.

\paraib{Data Science Analytics.}
This dataset can also be valuable for users or non-researchers interested in exploring blockchain data. It offers an opportunity for those who wish to better understand blockchains. 
For example, data scientists can utilize this dataset to explore and analyze blockchain data on public platforms such as Kaggle.
Data scientists widely use it to demonstrate their general skills in data analytics, machine learning, and data science. Therefore, this dataset can help data scientists acquire new skills, giving them a competitive edge in the market or improving their prospects of securing a job in a blockchain company.

We hope this dataset proves to be a valuable resource for research groups interested in conducting studies on \glspl{L2} blockchains and ZKsync Era. We plan to publish our dataset and code in a GitHub repository to facilitate this.

\section{Conclusion} \label{sec:conclusion}

% Main motivation behind this work
At a high level, this paper addresses a critical concern within the scientific community: \stress{the availability and accessibility of blockchain data}. Data-driven research is dependent on high-quality datasets to make meaningful findings. Although blockchain data are theoretically publicly available, practical challenges often prevent researchers, especially those without technical expertise, from obtaining and preprocessing them. In addition, the cost of using external services or deploying infrastructure to run archive nodes can be prohibitive.

% What we provide and did?
To address these challenges, we have collected, pre-processed, and made available our ZKsync dataset to facilitate access for researchers. We also provide a detailed background on blockchains, including blocks, transactions, receipts, and logs, designed to help non-experts understand and utilize the data effectively. To illustrate the potential of this dataset, we offer example analyses and discuss future research directions.

We believe our contribution will be valuable to researchers studying blockchain \gls{L2} ecosystems, particularly those interested in ZKsync. This dataset also adds to the existing body of \gls{L2} research. It is designed for ease of use, using the Python library Polars to allow straightforward data processing on a local laptop.

Finally, to promote scientific reproducibility and to support further research, we have made our dataset publicly available on GitHub~\cite{ZKsync-GitHub}.

%------------------------------------------------------------------------------

\bibliographystyle{splncs04}
\bibliography{references}

\begin{thebibliography}{10}
\providecommand{\url}[1]{\texttt{#1}}
\providecommand{\urlprefix}{URL }
\providecommand{\doi}[1]{https://doi.org/#1}

\bibitem{EIP-712}
{EIP-712: Typed structured data hashing and signing}. \url{https://eips.ethereum.org/EIPS/eip-712} (2017)

\bibitem{EIP-1559}
{EIP-1559: Fee market change for ETH 1.0 chain}. \url{https://eips.ethereum.org/EIPS/eip-1559} (2019)

\bibitem{EIP-2930}
{EIP-2930: Optional access lists}. \url{https://eips.ethereum.org/EIPS/eip-2930} (2020)

\bibitem{ZKsync-GitHub}
{Data sets and scripts used to analyze the ZKsync Era blockchain}. \url{https://github.com/matter-labs/zksync-data-dump} (2024)

\bibitem{ZKsync-doc-l2-to-l1-logs}
{L1 <-> L2 Communication}. \url{https://docs.zksync.io/zk-stack/concepts/l1_l2_communication} (2024)

\bibitem{Reth}
{Reth Book}. \url{https://reth.rs} (2024)

\bibitem{ZKsync-doc-contracts-bootloader}
{System contracts/bootloader description (VM v1.4.0)}. \url{https://github.com/code-423n4/2023-10-zksync/blob/main/docs/Smart contract Section/System contracts bootloader description.md} (2024)

\bibitem{ZKsync-doc}
{Transaction Lifecycle}. \url{https://docs.zksync.io/zk-stack/concepts/transaction-lifecycle\#transaction-types} (2024)

\bibitem{bagourd2023quantifying}
Bagourd, A., Francois, L.G.: Quantifying mev on layer 2 networks. arXiv preprint arXiv:2309.00629  (2023)

\bibitem{bertoni2009keccak}
Bertoni, G., Daemen, J., Peeters, M., Van~Assche, G.: Keccak sponge function family main document. Submission to NIST (Round 2)  (2009)

\bibitem{BitcoinCore-2024}
{bitcoin.org}: {Bitcoin Core}. \url{https://bitcoin.org/en/bitcoin-core} (2024)

\bibitem{BSC}
{bnbchain.org}: {BNB Smart Chain}. \url{https://www.bnbchain.org/en/bnb-smart-chain} (2024)

\bibitem{buterin_rollup-centric_2020}
Buterin, V.: A rollup-centric ethereum roadmap (Oct 2020), \url{https://ethereum-magicians.org/t/a-rollup-centric-ethereum-roadmap/4698}

\bibitem{day_introducing_2019}
Day, A., Medvedev, E., AK, N., Price, W.: Introducing six new cryptocurrencies in bigquery public datasets—and how to analyze them. Google Cloud  (2019)

\bibitem{Fan@WWW-Altruistic}
Fan, S., Min, T., Wu, X., Cai, W.: Altruistic and profit-oriented: Making sense of roles in web3 community from airdrop perspective. In: Proceedings of the 2023 CHI Conference on Human Factors in Computing Systems. CHI '23 (2023)

\bibitem{gogol2024liquid}
Gogol, K., Fritsch, R., Messias, J., Schlosser, M., Kraner, B., Tessone, C.: Liquid staking tokens in automated market makers  (2024)

\bibitem{gogol2024crossrollupmevnonatomicarbitrage}
Gogol, K., Messias, J., Miori, D., Tessone, C., Livshits, B.: Cross-rollup mev: Non-atomic arbitrage across l2 blockchains (2024)

\bibitem{gogol2024quantifying}
Gogol, K., Messias, J., Miori, D., Tessone, C., Livshits, B.: Quantifying arbitrage in automated market makers: An empirical study of ethereum zk rollups. arXiv preprint arXiv:2403.16083  (2024)

\bibitem{gogol2023cross}
Gogol, K., Messias, J., Schlosser, M., Kraner, B., Tessone, C.: Cross-border exchange of cbdcs using layer-2 blockchain. arXiv preprint arXiv:2312.16193  (2023)

\bibitem{Heimbach@IMC-PBS}
Heimbach, L., Kiffer, L., Ferreira~Torres, C., Wattenhofer, R.: Ethereum's proposer-builder separation: Promises and realities. In: Proceedings of the 2023 ACM on Internet Measurement Conference. IMC '23 (2023)

\bibitem{Heimbach-SP-Arbitrage}
Heimbach, L., Pahari, V., Schertenleib, E.: {Non-Atomic Arbitrage in Decentralized Finance}. In: 2024 IEEE Symposium on Security and Privacy (SP) (2024)

\bibitem{johnsson1984efficient}
Johnsson, T.: Efficient compilation of lazy evaluation. In: Proceedings of the 1984 SIGPLAN symposium on Compiler construction (1984)

\bibitem{L2beat}
{L2Beat}: {L2Beat: The state of the layer two ecosystem}. \url{https://l2beat.com/scaling/summary?sort-by=total&sort-order=desc\#layer2s} (2024)

\bibitem{Boojum}
{Matter Labs}: {Boojum Upgrade: zkSync Era’s New High-performance Proof System for Radical Decentralization}. \url{https://zksync.mirror.xyz/HJ2Pj45EJkRdt5Pau-ZXwkV2ctPx8qFL19STM5jdYhc} (2024)

\bibitem{Messias@IMC2021}
Messias, J., Alzayat, M., Chandrasekaran, B., Gummadi, K.P., Loiseau, P., Mislove, A.: Selfish \& opaque transaction ordering in the bitcoin blockchain: The case for chain neutrality. In: Proceedings of the 21st ACM Internet Measurement Conference. IMC '21 (2021)

\bibitem{messias2024writing}
Messias, J., Gogol, K., Silva, M.I., Livshits, B.: The writing is on the wall: Analyzing the boom of inscriptions and its impact on evm-compatible blockchains. arXiv preprint arXiv:2405.15288  (2024)

\bibitem{Messias@FC2023}
Messias, J., Pahari, V., Chandrasekaran, B., Gummadi, K.P., Loiseau, P.: {Dissecting Bitcoin and Ethereum Transactions: On the Lack of Transaction Contention and Prioritization Transparency in Blockchains}. In: Proceedings of the Financial Cryptography and Data Security (FC'23) (May 2023)

\bibitem{Messias-governance@ArXiv}
Messias, J., Pahari, V., Chandrasekaran, B., Gummadi, K.P., Loiseau, P.: Understanding blockchain governance: Analyzing decentralized voting to amend defi smart contracts (2024)

\bibitem{messias2023airdrops}
Messias, J., Yaish, A., Livshits, B.: Airdrops: Giving money away is harder than it seems. arXiv preprint arXiv:2312.02752  (2023)

\bibitem{Arbitrum}
{Offchain Labs}: {A gentle introduction to Arbitrum}. \url{https://docs.arbitrum.io/intro} (2024)

\bibitem{Omena@Poster}
Omena, J.J., Messias, J., Gouveia, F., Ventura, R.: Digital methods for blockchain research. \url{https://www.researchgate.net/publication/382511569_Digital_Methods_for_Blockchain_Research} (2024)

\bibitem{Godbole@Coindesk}
{Omkar Godbole}: {Layer 2 Blockchains Become Cheaper After Ethereum's Dencun Upgrade}. \url{https://www.coindesk.com/markets/2024/03/14/layer-2-blockchains-become-cheaper-after-ethereums-dencun-upgrade} (2024)

\bibitem{Optimism}
{Optimism Foundation}: {Optimism}. \url{https://www.optimism.io} (2024)

\bibitem{Pandas}
{Pandas}: {Pandas: Python data analysis library}. \url{https://pandas.pydata.org} (2024)

\bibitem{Paradigm-data-portal}
{Paradigm}: {Paradigm Data Portal}. \url{https://www.paradigm.xyz/oss/portal} (2024)

\bibitem{Paradigm-data-portal-git}
{Paradigm}: {Paradigm Data Portal}. \url{https://github.com/paradigmxyz/paradigm-data-portal} (2024)

\bibitem{Polars-Lazy}
{Polars}: {Lazy / Eager API: Polars user guide}. \url{https://docs.pola.rs/user-guide/concepts/lazy-vs-eager} (2024)

\bibitem{Polars}
{Polars}: {Polars: Dataframes for the new era}. \url{https://pola.rs} (2024)

\bibitem{qin2022quantifying}
Qin, K., Zhou, L., Gervais, A.: Quantifying blockchain extractable value: How dark is the forest? In: 2022 IEEE Symposium on Security and Privacy (SP) (2022)

\bibitem{qin2021attacking}
Qin, K., Zhou, L., Livshits, B., Gervais, A.: Attacking the defi ecosystem with flash loans for fun and profit. In: International conference on financial cryptography and data security (2021)

\bibitem{sharma2023unpacking}
Sharma, T., Potter, Y., Pongmala, K., Wang, H., Miller, A., Song, D., Wang, Y.: Unpacking how decentralized autonomous organizations (daos) work in practice. In: 2024 IEEE International Conference on Blockchain and Cryptocurrency (ICBC) (2024)

\bibitem{torres2021frontrunner}
Torres, C.F., Camino, R., State, R.: Frontrunner jones and the raiders of the dark forest: An empirical study of frontrunning on the ethereum blockchain. In: 30th USENIX Security Symposium (USENIX Security 21) (2021)

\bibitem{torres2024rolling}
Torres, C.F., Mamuti, A., Weintraub, B., Nita-Rotaru, C., Shinde, S.: Rolling in the shadows: Analyzing the extraction of mev across layer-2 rollups. arXiv preprint arXiv:2405.00138  (2024)

\bibitem{EIP-4844}
{Vitalik Buterin and Dankrad Feist and Diederik Loerakker and George Kadianakis and Matt Garnett and Mofi Taiwo and Ansgar Dietrichs}: {EIP-4844: Shard Blob Transactions}. \url{https://github.com/ethereum/EIPs/blob/master/EIPS/eip-4844.md} (2022)

\bibitem{weintraub2022flash}
Weintraub, B., Torres, C.F., Nita-Rotaru, C., State, R.: A flash (bot) in the pan: measuring maximal extractable value in private pools. In: Proceedings of the 22nd ACM Internet Measurement Conference (2022)

\end{thebibliography}

% Add the appendix in a pre-print version
\appendix
\section{Additional Information for Data Schemas} \label{sec:appendix}

In this section, we present the attributes available in our dataset, including their data types and brief descriptions. Our dataset consists of one year of ZKsync data, containing all blocks, transactions, receipts, and logs included in the ZKsync blockchain.

Table~\ref{tab:block_attributes} describes the attributes of each block added to the ZKsync blockchain. Table~\ref{tab:transaction_attributes} details the transaction data, including their data types and descriptions. Table~\ref{tab:transaction_receipts_attributes} provides information on transaction receipts, which are generated immediately after a transaction is executed and added to a block. This is useful for computing the actual gas spent on a transaction and the corresponding transaction fees paid by users.

Next, Table~\ref{tab:log_attributes} contains information about transaction logs, which is crucial for analyzing changes in smart contract states such as account balances, token transfers, votes cast, and swaps. Finally, Table~\ref{tab:log_l2_l1_attributes} presents the attributes related to the messages emitted by ZKsync to the Ethereum mainnet.

\begin{table*}[tb]
\centering
\resizebox{\textwidth}{!}{%
\begin{tabular}{rrp{9cm}}
\toprule
\thead{Attribute} & \thead{Type} & \thead{Description} \\
\midrule
hash & str & Unique identifier for the block. \\
parentHash & str & Unique identifier of the parent block. \\ 
sha3Uncles & str & SHA-3 hash of the uncles' block headers. On ZKsync it is set to \stress{0x1dcc $\cdots$ 9347} since there are no uncle blocks. \\
miner & str & Address of the miner who mined the block. This is set as Null address (\stress{0x0}) in all ZKsync blocks since it does not have miners or block validators. \\
stateRoot & str & Root hash of the state trie. Set to Null address (\stress{0x0}). \\
transactionsRoot & str & Root hash of the transaction trie. Set to Null address (\stress{0x0}). \\
receiptsRoot & str & Root hash of the receipts trie. Set to Null address (\stress{0x0}). \\
number & i64 & Block number or height. \\
l1BatchNumber & str & L1 batch number, related to the sequence of batches submitted to Layer 1 on zkRollup systems. \\
gasUsed & i64 & Total amount of gas used by all transactions in the block. \\
gasLimit & i64 & Maximum amount of gas that can be used by all transactions in the block. Set to \num{4294967296} ($2^{32}$) units of gas. \\
baseFeePerGas & i64 & Base fee, in Wei ($10^{-18} ETH$), per unit of gas. \\
extraData & str & Extra data included by the miner in the block. Set to \stress{0x} since there are no miners. \\
logsBloom & str & Bloom filter for the logs of the block. Set to \stress{0x0 $\cdots$ 0}. \\
timestamp & i64 & Unix timestamp of when the block was collated. \\
l1BatchTimestamp & str & L1 batch timestamp (in HEX format) associated with the block. \\
difficulty & i64 & Difficulty target for mining the block. Set to 0 since there is no mining. \\
totalDifficulty & i64 & Cumulative difficulty of the blockchain up to and including this block. Set to 0 since there is no mining. \\
sealFields & list[null] & Seal fields containing proof-of-work or proof-of-stake information. List containing null as value. \\
uncles & list[null] & Uncle blocks that were mined but not included in the main chain. List with Null value since there is no mining. \\
size & i64 & Size of the block in bytes. Set to 0. \\
mixHash & str & Hash used in the mining process to prove that enough computational work has been performed. Set to \stress{0x0 $\cdots$ 0} since there is no mining. \\
nonce & str & Value used in the mining process to find a valid block hash. Set to \stress{0x0 $\cdots$ 0} since there is no mining. \\
\bottomrule
\end{tabular}}
\caption{Description of block attributes on ZKsync.}
\label{tab:block_attributes}

\end{table*}

\begin{table*}[tb]
\centering
\resizebox{\textwidth}{!}{%
\begin{tabular}{rrp{9cm}}
\toprule
\thead{Attribute} & \thead{Type} & \thead{Description} \\
\midrule
blockHash & str & Unique identifier of the block containing the transaction. \\
blockNumber & i64 & Block number or height containing the transaction. \\
chainId & i64 & Identifier of the blockchain network. Set to 324 that represents ZKsync chain. \\
from & str & Address of the sender of the transaction. \\
gas & i64 & Amount of gas provided as \stress{gasLimit} for the transaction. \\
gasPrice & i64 & Price per unit of gas the sender is willing to pay. \\
hash & str & Unique identifier for the transaction. \\
input & str & Data sent along with the transaction. In HEX code. \\
l1BatchNumber & str & L1 batch number related to the transaction in zkRollup systems. \\
l1BatchTxIndex & str & Index of the transaction in the L1 batch. \\
maxFeePerGas & i64 & Maximum fee, in Wei ($10^{-18}$ ETH), per unit of gas. \\
maxPriorityFeePerGas & i64 & Maximum priority fee, in Wei ($10^{-18}$ ETH), per unit of gas. \\
nonce & i64 & Number of transactions sent by the sender prior to this one. \\
r & str & First part of the ECDSA signature. \\
s & str & Second part of the ECDSA signature. \\
to & str & Address of the receiver of the transaction. \\
transactionIndex & i64 & Index of the transaction within the block. \\
type & i64 & Type of transaction divided into 5 categories: \stress{Legacy} (0 or \stress{0x0}), \stress{EIP-2930} (1 or \stress{0x1})~\cite{EIP-2930}, \stress{EIP-1559} (2 or \stress{0x2})~\cite{EIP-1559}, \stress{EIP-712} (113 or \stress{0x71})~\cite{EIP-712}, and \stress{Priority} (255 or \stress{0xff}). See details in ZKsync documentation~\cite{ZKsync-doc}. \\
v & f64 & Recovery id of the ECDSA signature. \\
value & str & Amount of tokens (in Wei and in HEX format) that are transferred in the transaction to the recipient address (\stress{to}). \\
\bottomrule
\end{tabular}}
\caption{Description of transaction attributes on ZKsync.}
\label{tab:transaction_attributes}
\end{table*}

\begin{table*}[tb]
\centering
\resizebox{\textwidth}{!}{%
\begin{tabular}{rrp{9cm}}
\toprule
\textbf{Attribute} & \textbf{Type} & \textbf{Description} \\
\midrule
blockHash          & str   & Unique identifier of the block containing the transaction. \\
blockNumber        & i64   & Block number or height containing the transaction. \\
contractAddress    & str   & Address of the contract created by the transaction, if applicable. \\
cumulativeGasUsed  & i64   & Total amount of gas used when the transaction was executed in the block. Set to 0 on ZKsync. \\
effectiveGasPrice  & i64   & Actual price per unit of gas, in Wei ($10^{-18}$ ETH), paid. \\
from               & str   & Address of the sender of the transaction. \\
gasUsed            & i64   & Amount of gas used by the transaction. \\
l1BatchNumber      & str   & L1 batch number related to the transaction in zkRollup systems. \\
l1BatchTxIndex     & str   & Index of the transaction in the L1 batch. \\
logsBloom          & str   & Bloom filter for the logs of the transaction. Set to \stress{0x0 $\cdots$ 0} on ZKsync. \\
root               & str   & State root after the transaction is executed. \\
status             & i64   & Status of the transaction (1 for success, 0 for failure). \\
to                 & str   & Address of the receiver of the transaction. \\
transactionHash    & str   & Unique identifier for the transaction. \\
transactionIndex   & i64   & Index of the transaction within the block. \\
type               & i64   & Type of transaction divided into 5 categories. Refer to Table~\ref{tab:transaction_attributes} and to ZKsync documentation~\cite{ZKsync-doc} for details.\\
\bottomrule
\end{tabular}}
\caption{Description of transaction receipt attributes on ZKsync.}
\label{tab:transaction_receipts_attributes}
\end{table*}

\begin{table*}[tb]
\centering
\resizebox{\textwidth}{!}{%
\begin{tabular}{rrp{9cm}}
\toprule
\textbf{Attribute} & \textbf{Type} & \textbf{Description} \\
\midrule
address               & str   & Address of the contract that generated the log. \\
blockHash             & str   & Unique identifier of the block containing the transaction. \\
blockNumber           & i64   & Block number or height containing the transaction. \\
data                  & str   & Data contained in the log. This can be used, for example, to extract the amount of tokens transferred from one user to another. \\
l1BatchNumber         & str   & L1 batch number related to the log in zkRollup systems. \\
logIndex              & i64   & Index of the log within the block. \\
logType               & null  & Type of log. Set to \stress{null} on ZKsync. \\
removed               & bool  & Indicates whether the log was removed (\stress{true}) or not (\stress{false}). \\
transactionHash       & str   & Unique identifier for the transaction. \\
transactionIndex      & i64   & Index of the transaction within the block. \\
transactionLogIndex   & str   & Index of the log within the transaction. In HEX format. \\
topics$_0$              & str   & First topic of the log. This is typically referred to as the name of the event encoded in hexadecimal (HEX) format. \\
topics$_1$             & str   & Second topic of the log. \\
topics$_{2}$             & str   & Third topic of the log. \\
topics$_3$              & str   & Fourth topic of the log. \\
\bottomrule
\end{tabular}}
\caption{Description of transaction log attributes on ZKsync.}
\label{tab:log_attributes}
\end{table*}

\begin{table*}[tb]
\centering
\resizebox{\textwidth}{!}{%
\begin{tabular}{llp{9cm}}
\toprule
\textbf{Attribute} & \textbf{Type} & \textbf{Description} \\
\midrule
blockHash           & str   & Unique identifier of the block containing the log. \\
blockNumber         & str   & Block number or height containing the log. \\
isService           & bool  & Indicates whether the log is a service log (\stress{true}) or not (\stress{false}). \\
key                 & str   & Key associated with the log that could be used to carry some data with the log. \\
l1BatchNumber       & str   & L1 batch number related to the log in zkRollup systems. \\
logIndex            & str   & Index of the log within the block. \\
sender              & str   & It is the value of \stress{this} in the frame where the L2→L1 log was emitted. \\
shardId             & str   & It is the id of the shard the opcode was called. It is currently set to 0. \\
transactionHash     & str   & Unique identifier for the transaction associated with the log. \\
transactionIndex    & str   & Index of the transaction within the block. \\
transactionLogIndex & str   & Index of the log within the transaction. \\
txIndexInL1Batch    & str   & Index of the transaction within the L1 batch. \\
value               & str   & Value associated with the log that could be used to carry some data with the log. \\
\bottomrule
\end{tabular}}
\caption{Description of L2 to L1 log attributes on ZKsync.}
\label{tab:log_l2_l1_attributes}
\end{table*}

\section{Glossary}
\label{sec:glossary}
Following is a list of important notations used in this paper.
% START Make sure that the symbols and acronyms have constant indentation
% \renewcommand{\glossarysection}[2][]{}  % Disables the titles
\setglossarystyle{alttree}
\glssetwidest{AAAA}
% END
% \printnoidxglossary[type={symbols}]
\printnoidxglossary[type={acronym}]

\end{document}